\documentclass[preprint,aps,letter,superscriptaddress,floatfix]{revtex4}

\usepackage{graphicx}
\usepackage[ansinew]{inputenc}
\usepackage{array}
\usepackage{color}
\usepackage{psfrag}
\usepackage{amsmath}
\usepackage{amsxtra}
\usepackage{amstext}
\usepackage{amssymb}

\usepackage{latexsym}
\usepackage{dsfont}

\newcommand{\bi}{\bibitem}

\newcommand{\be}{\begin{equation}}
\newcommand{\ee}{\end{equation}}

\begin{document}
\renewcommand{\figurename}{Figure}

\title{Suppression of Interactions in Multimode Random Lasers in the Anderson Localized Regime}
\author{Peter Stano}
\affiliation{Department of Physics, University of Basel, Klingelbergstrasse 82, CH-4056 Basel, Switzerland}

\author{Philippe Jacquod}
\affiliation{Physics Department, University of Arizona, 1118 E 4th Street, Tucson, AZ 85721, USA}
\affiliation{College of Optical Sciences, University of Arizona, Tucson, AZ 85721, USA}
\affiliation{D\'epartment de Physique Th\'eorique, Universit\'e de Gen\`eve CH-1211 Gen\`eve, Switzerland}

\date{\today}

\begin{abstract}
Understanding random lasing is a formidable theoretical challenge. 
Unlike conventional lasers,
random lasers have no resonator to trap light, they are highly multimode with potentially strong modal
interactions and they are based on disordered gain media, where photons
undergo random multiple scattering~\cite{Cao99,Fro99,Wie08}. Interference effects notoriously
modify the 
propagation of waves in such random media, but their fate in the presence of nonlinearity
and interactions is poorly understood. Here, we present a semiclassical theory for multimode
random lasing in the strongly scattering regime. 
We show that Anderson localization~\cite{And58,Lag09}, a wave-interference effect,
is not affected by the presence of nonlinearities. To the contrary, its presence
suppresses interactions between simultaneously lasing modes.
Using a recently 
constructed theory for complex multimode lasers~\cite{Tur}, we show analytically 
how Anderson localization 
justifies a noninteracting, single-pole approximation. 
Consequently, lasing modes in a strongly scattering random laser 
are given by long-lived, Anderson localized modes of the passive cavity,
whose frequency and wave profile does not vary with pumping, even 
in the multi-mode regime when mode overlap spatially.
\end{abstract}


\maketitle

Wave interference effects influence a broad range of physical processes
in wave propagation through random media~\cite{Ish78,Akk07}, ranging
from the time-evolution of ocean waves and the transmission of light through interstellar clouds, down to
the dynamics of individual electrons in nano-electronic circuits and the transmission of photons through
imperfect optical fibers and other waveguides. They generate a rich variety of wave-interference 
phenomena such as Anderson localization~\cite{And58,Lag09} and
coherent backscattering~\cite{Backsc1,Backsc2,Backsc3}, as well as
effects of higher-order coherence such as quantum noise and particle bunching
or anti-bunching~\cite{coherence}. 
For linear wave equations and noninteracting fields, these effects are well understood~\cite{Akk07},
however little is known of the influence that nonlinearities and interactions have on
coherent wave interference effects.
In random lasers, confinement of light within an amplifying media is achieved via multiple
scattering~\cite{Cao99,Fro99,Wie08}
and one thus expects important wave-interference effects
to be present. 
The mechanism behind the observed spectral narrowing has been and still is a matter of debate,
with most theoretical proposals constructing lasing modes from modes of the passive cavity
trapped by scattering, be they
Anderson localized modes in the strong scattering regime~\cite{Jia00,Van01} or 
anomalously trapped modes in the weak scattering regime~\cite{Apa04}. 
Numerical calculations suggest that lasing in weakly scattering 
random lasers can be supported by extended
diffusive modes~\cite{Van07,Tur08}. 
Recent experiments have observed delocalized lasing modes in weakly scattering
random lasers~\cite{Muj04,Tul10}, while lasing modes were significantly less extended for stronger
scattering~\cite{Mol07}, with some evidence for mode-dependent 
localization lengths~\cite{Fal09}. The Anderson localized regime has been probed 
in few experiments,
where  the spatial extent of lasing modes could not be directly accessed~\cite{Mil05}.
Because random lasers mostly operate in the multiple scattering, multi-mode regime,
they are perfect testbeds to investigate effects of wave coherence in the presence of nonlinearities and
interactions.
With the exception of Ref.~\cite{Tur08}, theoretical and numerical
investigations of the localization properties of lasing modes rely on
calculations of the total electromagnetic field inside the amplifying medium.
There is still very little understanding of the localization 
properties of individual
lasing modes, how they are influenced by the infinite-order nonlinearity arising from
spatial-hole burning and by the resulting modal interactions.
Of particular interest is to find out if, above lasing threshold,
modal interactions change the localization length in an Anderson localized passive cavity.
The influence of modal interactions usually becomes more and more pronounced as
more modes start to lase, thus it is expectable that lasing modes differ more and more from 
passive cavity modes as the pump strength increases. 

We focus on the interplay between Anderson localization and nonlinearities
in a multi-mode random laser, which we take as a photon-scattering
gain medium region surrounded by vacuum. 
Below threshold, the system is linear and we assume the gain medium to be strongly scattering, 
so that the modes of the passive cavity are exponentially localized over a localization
length $\xi$  much smaller than the cavity's linear size, $\xi \ll L$. This simultaneously means that
modes are long-lived,
with a mean lifetime $\tau_\xi \simeq (\xi/c) \exp[L/\xi]$ and low lasing thresholds when they are
localized deep enough inside the 
cavity~\cite{Sta00}. 
Our goal is to understand how lasing modes individually differ from passive cavity modes. 
To that end we adopt the 
method of Ref.~\cite{Tur} (see Supplementary Information),
where the electric field is expanded
as  $E(x,t) = \sum_\mu \Psi_\mu(x) \exp[-i \Omega_\mu t]$ over individual lasing modes $\Psi_\mu$. The latter 
are in their turn expanded 
over so-called constant flux (CF) modes, which are eigenmodes  of the passive cavity
satisfying the boundary condition that they couple only to undamped outgoing waves
(i.e. with real wavenumber $k_{\mu}$) outside the cavity.
The resulting modal
resolution of $E(x)$ allows to directly compare individual lasing modes
with the CF modes. Our main finding is that, quite surprisingly, simultaneously lasing modes
in the Anderson localized regime do not interact with each other. 
To a very high precision, the frequency and wavefunction
of a lasing mode are the same as that of a
CF state,  regardless of the pump strength, even in the multi-mode regime and as new modes
start to lase. Nonlinearities only decide which mode starts to lase at what pump 
threshold. Our analysis of the lasing
theory of Ref.~\cite{Tur}, augmented by wavefunction correlations in the Anderson localized
regime~\cite{Ivanov}, provides a straightforward analytical understanding of the suppression 
of modal interactions: a single-pole approximation is rigorously  justified, 
even in the multi-mode regime,
because of the exponentially small broadening of  localized modes and their exponentially
small coupling to only an algebraic number of CF modes.
The absence of modal interaction we predict is totally unexpected: 
a similar system, with the same gain medium and cavity shape and size, but without or with little
scattering standardly exhibits strong mode competition, with in particular 
shifts in mode frequencies and the disappearance (and possible re-appearance) of certain modes
as the pump strength increases~\cite{Tur08}.
Modal interactions are turned on once CF modes are broadened, which can occur
either at weaker disorder when their localization length increases, or when their lifetime
is limited by photon absorption.
The linear noninteracting behavior of lasing modes we predict is the trademark of
Anderson localization in random lasers, and we propose to monitor
lasing frequencies as a function of
pump strength to detect Anderson localization in experiments on random lasers.

As starting point we take the Maxwell-Bloch equations (MBE), a set of three coupled nonlinear
equations for the electric field, the polarization and the population inversion in the
gain medium~\cite{Haken} (see Supplementary Information).  For simplicity we consider either
a one-dimensional system or a two-dimensional system with electromagnetic 
field transverse electric with polarization perpendicular to the gain medium, or transverse
magnetic. In those instances, it is sufficient to consider scalar fields.
To find steady-state solutions, Ref.~\cite{Tur} first writes  both 
the polarization and the electric field  in a multiperiodic expansion. Further imposing
the rotating wave and stationary inversion approximations, one obtains
a set of self-consistent equations
\be\label{eq:map}
{\bf a}_\mu = D_0 {\bf T}\, {\bf a}_\mu \equiv D_0 \Lambda \mathcal{T} {\bf a}_\mu \, ,
\ee
for the vector ${\bf a}_\mu = (a^1_\mu,a^2_\mu, ... , )$ of
components of the expansion (we use dimensionless units for
$\Psi$ and $D_0$, as described in the Supplementary Information)
\be \label{eq:expansion}
\Psi_\mu({\bf x},k_\mu) = \sum_{m=1}^{M_\mu} a_\mu^m \psi_m({\bf x},k_\mu) \, , 
\ee
of the lasing modes $\Psi_\mu({\bf x},k_\mu)$ over $M_\mu$ elements of a truncated 
basis of CF modes $\{ \psi_m({\bf x},k_\mu) \}_{m=1}^M$. When $N$ modes are lasing,
$M \geq N$, and
in the Anderson localized regime, 
we will see that the structure of the threshold matrix ${\bf T}$  is such that
$M_\mu=1$ for each lasing state. In Eq.~(\ref{eq:map}), 
$D_0$ gives the overall pump strength, which we define as the population inversion density of the active medium.
The threshold matrix 
${\bf T} \equiv  \Lambda {\cal T}$ is the product of a diagonal and a non-diagonal matrix,
\begin{subequations}\label{eq:mmnn}
\begin{eqnarray}\label{eq:mm}
\Lambda_{mn} &=& \delta_{mn} \frac{-k_\mu^2}{k_\mu^2-k_{m\mu}^2}\frac{\gamma_\perp}{ck_\mu-\omega_0+{\rm i}\gamma_\perp} \, ,\\
\label{eq:mn}
\mathcal{T}_{mn} &=& \int_{\cal C} {\rm d}^dx 
\frac{F({\bf x})}{1+h({\bf x})} \psi_m({\bf x},k_\mu) \psi_n({\bf x},k_\mu) \, ,
\end{eqnarray}
\end{subequations} 
with the pump profile $F({\bf x})$, the speed of light $c$, the atomic frequency $\omega_0$ of the two-level 
active medium and the polarization relaxation rate $\gamma_\perp$. 
The integral runs over the volume ${\cal C}$ (length in dimension $d=1$, area in $d=2$)  of the gain medium.
Nonlinearities arise from the hole burning denominator
\be\label{eq:holeb}
h({\bf x}) = \sum_\nu \Gamma(k_\nu) |\Psi_\nu({\bf x})|^2 \, ,
\ee
where the sum runs over all lasing modes and 
$\Gamma(k) = \gamma_\perp^2 / [ (c k - \omega_0 )^2 + \gamma_\perp^2 ]$.
The lasing modes are self-consistently determined by the nonlinear eigenvalue problem of
Eq.~(\ref{eq:map}). Solutions exist only for discrete real values of $k_\mu$, giving the lasing frequencies
$\Omega_\mu = c k_\mu$. The wavenumber of the 
CF modes $k_{m \mu} = \overline{k}_{m \mu} + i 
\kappa_{m \mu}$, with $\overline{k}_{m \mu},
\kappa_{m \mu} \in {\cal R}$,
are complex because of the openness of the cavity, $\kappa_{m \mu}^{-1} = c \tau_{m \mu}$ with 
the lifetime $\tau_{m \mu}$.

The presence of modal interactions is reflected in the structure of the lasing modes, e.g. the number 
$M_\mu$ of CF states significantly contributing to a single lasing mode in the expansion of Eq.~(\ref{eq:expansion}). 
If the expansion is dominated by a single
component, the lasing mode resembles a noninteracting CF mode, however if the expansion runs
over many nonvanishing 
components, the lasing mode is in general very different from noninteracting CF states, 
signaling the onset of modal interactions.
We next proceed to show that $M_\mu \simeq 1$ in the strongly localized regime,
by analyzing the structure of the  
threshold matrix  ${\bf T} = \Lambda {\cal T}$. In the Supplementary Information,
we argue that matrix elements ${\cal T}_{mn}$ are non-zero only when CF states $\psi_m$ and
$\psi_n$ are centered within one localization length of one another. Furthermore, 
the nonvanishing matrix elements
fluctuate pseudo-randomly as a function of the indices 
$(m,n)$. 
In the Anderson localized regime, most
CF modes are long-lived,  $\kappa_{m \mu} \simeq \xi^{-1} \exp[-L/\xi]$ and
consequently  $|\Lambda_{mm}| \simeq 
(k_\mu \xi \, \gamma_\perp/2) \exp[L/\xi] \, \big/(c k_\mu-\omega_0+{\rm i} \gamma_\perp)$ is exponentially large
at $k_\mu = \overline{k}_{m \mu}$. One then expects that lasing frequencies 
are very close to those of certain CF modes,
$k_\mu \simeq \overline{k}_{m_0 \mu}$, and that the expansion (\ref{eq:expansion}) for the
corresponding lasing mode is dominated by a single component $a_\mu^{m_0}$. 
We show perturbatively
that this is indeed the case
as long as $\exp[-L/\xi] \ll 1$. Keeping only a single component in
Eq.~(\ref{eq:map}) gives the single-pole approximation, $(a_\mu^n)^{(0)}=0$ for $n\neq m_0$, and
\begin{eqnarray}\label{eq:spole}
(a_\mu^{m_0})^{(0)} = D_0 {\bf T}_{m_0 m_0} \,  (a_\mu^{m_0})^{(0)} \, .
\end{eqnarray}
At this level, one has $D_0 = {\bf T}_{m_0 m_0}^{-1} = (\Lambda_{m_0 m_0}
{\cal T}_{m_0 m_0})^{-1} \propto \exp[-L/\xi] \ll 1$.
Leading-order corrections are obtained with the approximation 
\begin{eqnarray}\label{eq:first_a}
(a_\mu^{m_0})^{(1)} & =&  D_0 {\bf T}_{m_0 m_0} \,  (a_\mu^{m_0})^{(1)} + D_0^2 
\sum_{n \ne m_0} {\bf T}_{m_0 n} {\bf T}_{n m_0} (a_\mu^{m_0})^{(1)} \, .
\end{eqnarray}
From Eqs.~(\ref{eq:spole}) and (\ref{eq:first_a}), we conclude that 
corrections to the single-pole approximation are negligible when 
\begin{equation}\label{eq:cond}
 \Xi \equiv \frac{ \sum_{n \ne m_0} \Lambda_{nn}  {\cal T}_{m_0 n} {\cal T}_{n m_0} }{
\Lambda_{m_0 m_0}  {\cal T}_{m_0 m_0}^2 }
 \ll 1 \, ,
\end{equation}
with $\Lambda_{mn}$ and ${\cal T}_{mn}$ defined in Eqs.(\ref{eq:mmnn}). 
We next analyze the conditions under which Eq.~(\ref{eq:cond}) is satisfied.

In order for ${\cal T}_{m_0 n} {\cal T}_{n m_0}$ not to vanish, the CF
states $m_0$ and $n$ must have their support within one localization length. At energy
$\hbar c k_\mu$ they must
thus differ in energy  by at least the CF mode spacing within one localization
volume~\cite{Ivanov}, in our case $\Delta_\xi = \hbar c/b_d k_\mu^{d-1} \xi^d$, $b_1=2$ and $b_2 = 2 \pi$. 
We have the upper
bound $|\Lambda_{n n}| \lesssim \hbar^2 c^2 k^2_\mu/(2 \hbar c k_\mu \Delta_\xi + \Delta_\xi^2) 
\times |\gamma_\perp/(c k_\mu -\omega_0+i \gamma_\perp)|$ for $n \ne m_0$ in 
Eq.~(\ref{eq:cond}), and thus
$|\Lambda_{n n} /  \Lambda_{m_0 m_0}| \lesssim 
2 b^2_d (k_\mu \xi)^{2d-1} \exp[-L/\xi] \big/ [1+2 b_d (k_\mu \xi)^d]$. Next, 
averages 
$\overline{{\cal T}_{m_0n}{\cal T}_{nm_0}}$ of product of matrix elements of ${\cal T}$
can be evaluated, under the assumption that 
CF states are sufficiently close to localized modes of a disordered closed system. This assumption
is satisfied when $\xi \ll L$. We obtain
(see Supplementary Information)
$\langle |\Xi| \rangle  \lesssim  f(k_\mu \xi) \exp[-L/\xi]$, with $f(x)$ an algebraic function of $x$.
Furthermore, it can also be shown that fluctuations of $\Xi$ are also exponentially small,
$\propto \exp[-L/\xi]$.
This gives an exponentially small upper bound for corrections to the single-pole approximation, 
both for $d=1$ ($\xi = \ell$) and
for $d=2$ when $\xi$ is not exponentially larger than the wavelength $2 \pi/k_\mu$. 
We conclude that the condition expressed in Eq.~(\ref{eq:cond}) 
is satisfied and that
a single-pole approximation is 
justified in the Anderson localized regime. Accordingly, we propose that the
criteria for the absence of
interactions in random lasers are $\xi \ll L$ in $d=1$ and $\xi \ll L$, $k_\mu \ell  \lesssim L/\xi$ in $d=2$.
While we expect that the $d=3$ criterion is the same as in two dimensions, we recall that 
an analysis of the MBE for vector fields is required in this case.

Corrections to the single-pole approximation become non-negligible when the mode lifetime
becomes shorter. We briefly discuss two cases when this happens.
First, in the Anderson localized regime, CF modes have an exponentially long lifetime, 
but their broadening can be limited by photon absorption. This can be
included in the above estimate via the substitution $\kappa_{m \mu} \simeq \xi^{-1} \exp[-L/\xi]
\rightarrow \kappa_{m \mu} \simeq (c \tau_{\rm abs})^{-1}$. This leads to 
$|\Lambda_{n n} /  \Lambda_{m_0 m_0}|_{\rm abs} \lesssim \hbar/ \Delta_\xi \tau_{\rm abs}$
and corrections to the single-pole approximations become
of the order 
$\Xi \sim  f(k_\mu \xi)/(k_\mu c \tau_{\rm abs})$.
We conclude that photon absorption turns on modal interactions in the localized regime, 
when it reduces the mode lifetime such that
$\tau_{\rm abs} \lesssim f(k_\mu \xi) \big/ k_\mu c$. 
Second, mode lifetimes are shorter and
wavefunction correlations are different in the $d=2$ diffusive, delocalized regime, where all
CF modes overlap. From Ref.~\cite{Mirlin}, we estimate (see Supplementary Information)
$\Xi \sim  \hbar f(k_\mu \ell,k_\mu L)  \big/ \Delta_L \tau_{m_0 \mu}$,
with the level spacing $\Delta_L = \hbar c\big/2 \pi k_\mu L^2$. Prelocalized states exist in such systems~\cite{Apa04},
whose lifetimes are long enough, $\tau_{\rm preloc} \gtrsim \hbar/\Delta_L $ that deviations from 
the single-pole approximation are still negligible 
if $ f(k_\mu \ell,k_\mu L) \ll \Delta_L \tau_{\rm preloc}/\hbar$. 
Ref.~\cite{Tur08} on the other hand suggests that lasing in diffusive random lasers can be supported by
standard generic diffusive 
modes. The lifetime of such modes can be estimated by the classical diffusion 
time $\tau_{\rm D}= L^2/c \ell$ through the system, and the dimensionless
measure of corrections to the single-pole approximation becomes
$\Xi \sim  k_{\mu} \ell  \times f(k_\mu \ell,k_\mu L)  > 1$. This agrees with the observed strong modal interactions reported in 
Ref.~\cite{Tur08} for a diffusive laser with $k_\mu \ell \gg 1$.
We caution however that our estimates all rely on 
the assumption that CF modes have the same wavefunction correlations as modes
of closed, nonabsorbing Anderson localized systems. This is no longer true
in the weakly scattering, delocalized regime.

We numerically check our theoretical predictions. From the above analysis, dimensionality
plays only a limited role in the Anderson localized regime, and for numerical
convenience, we take a $1d$ edge-emitting, disordered microcavity laser. 
The cavity extends from $x=-L/2$, where we impose reflecting boundary
conditions to $x=L/2$ where it is open. We fix its length, $L=2\, \mu$m and
the resonant frequency $\omega_0=c k_0$ with $k_0 L=50$.
The transversal and longitudinal relaxation rates are $\gamma_\perp=8 \times 10^{-2}\omega_0$ and $\gamma_{||}=2\times 10^{-5}\omega_0$.
Disorder is introduced 
in the cavity by a spatially varying refractive index randomly distributed in the interval
$n(x) \in [1.3,6.7]$ with a spatial correlation $\chi=8$ nm. This rather large variation
has been chosen to ensure that with the numerically reachable values of $L$ and $k_0$,
we are in the Anderson localized regime.
The localization length of cavity modes is numerically estimated with the inverse participation ratio
\be
{\rm I}(\Psi) = \left| \int {\rm d}x |\Psi(x)|^2 \right|^2 \Big/ \int {\rm d}x |\Psi(x)|^4 \, ,
\label{eq:IPR}
\ee 
which estimates the spatial extent of a wavefunction regardless of its overall normalization.
For an exponentially localized
mode $\Psi(x) = \exp[i \theta(x)] \exp[-|x|/2 \xi]/\sqrt{2 \xi}$,  one has ${\rm I}(\Psi) =4 \xi$.

\begin{figure}[t]
\includegraphics[width=16cm]{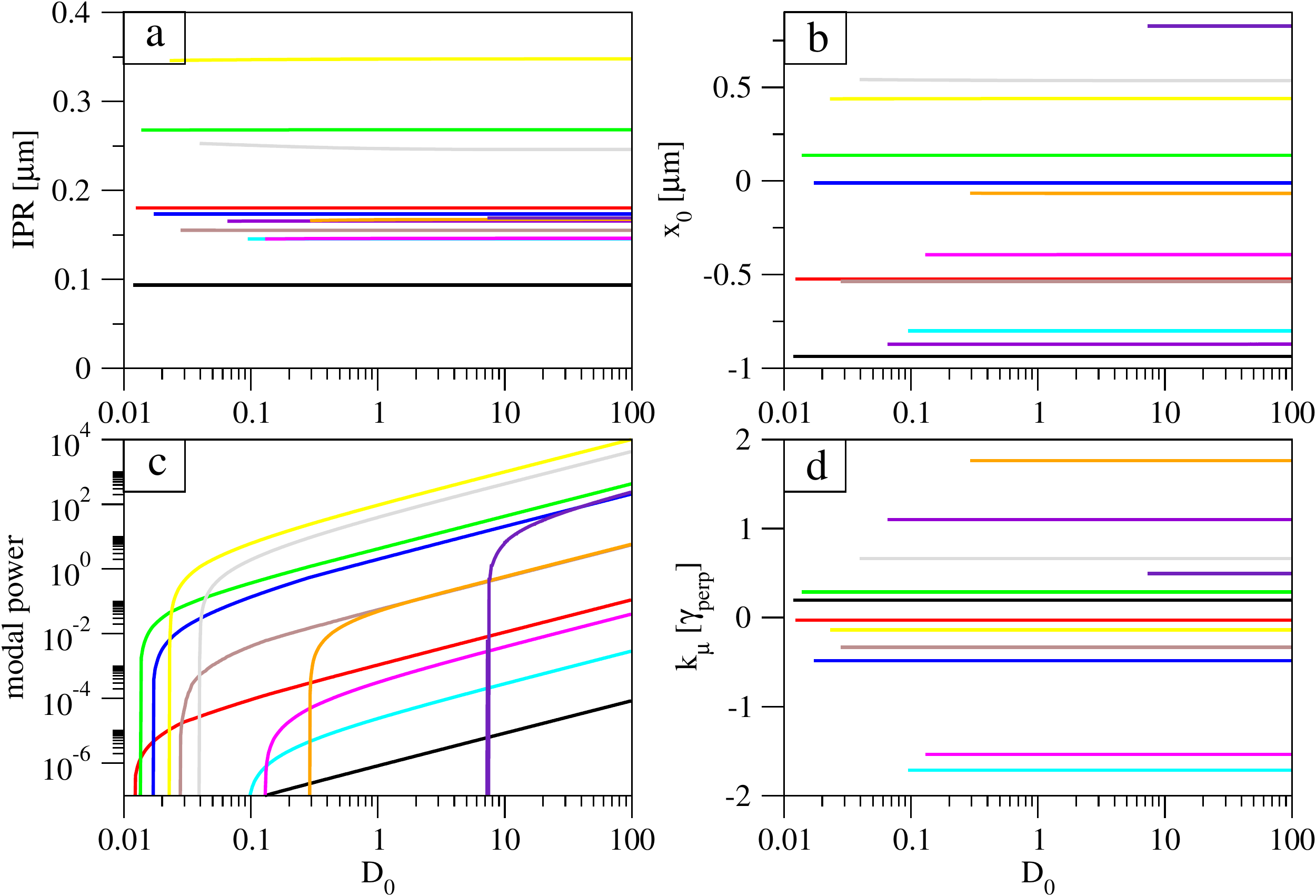} 
\caption{
Evolution of lasing modes as a function of pump strength $D_0$ for
a non-absorbing, Anderson localized, half-open $1d$ microcavity laser. 
The index of refraction is randomly distributed, $n({\bf x}) \in [1.3,6.7]$, 
with a correlation length $\chi=8$ nm. 
In each
panel, lines appear when the corresponding mode starts to lase.
Panel a): Inverse participation ratio defined in Eq.~(\ref{eq:IPR}), giving a measure of 
the spatial extent of the mode.
Panel b): Position of the center of mass 
$x_0 = \int {\rm d}x \, x |\Psi_\mu(x)|^2 /\int {\rm d}x \, |\Psi_\mu(x)|^2$ of the mode.
Panel c): Modal output power. The output power does not saturate because we did not limit
the population inversion density
in our numerics in order to investigate multi-mode lasing.
Panel d): Lasing frequency in units of the population inversion relaxation rate $\gamma_\perp$.}
\label{fig:localized}
\end{figure}

In Fig.~\ref{fig:localized} we show a typical set of data for lasing modes in the strongly
disordered, localized regime, indicating the mode's
position, spatial extent, lasing frequency and output power.
We estimate $\xi \simeq  0.03-0.1 \, \mu$m from Fig.~\ref{fig:localized}a,
so that $\xi/L \simeq 0.015-0.05 \ll 1$ and we are in the Anderson localized regime.
As usual, localization lengths vary from one mode to another. Lasing modes have long lifetimes,
and are localized well inside the cavity, a distance larger than $\xi$ away from the emitting
cavity boundary. 
We see that as the pump
varies by a factor of a thousand and up to twelve modes lase, lasing modes change neither their
position (monitored by their center of mass, Fig.~\ref{fig:localized}b), nor their frequency
(Fig.~\ref{fig:localized}d), nor their spatial extent (Fig.~\ref{fig:localized}a). This is even more remarkable,
given that several modes strongly overlap either spatially (for instance the red and brown modes)
or in frequency. Only the
modal emission power increases with pump strength, with a linear dependence on $D_0$
that is reached quickly after threshold for each lasing mode. This is what is expected for noninteracting
modes, given that $D_0$ gives the inversion density of the active medium and is not bound in 
our approach. Incorporating a finite maximal density in our model would lead to mode
saturation, but would also limit the number of lasing modes and thus constrain our 
investigations of mode interactions. 

\begin{figure}[t]
\includegraphics[height=5cm]{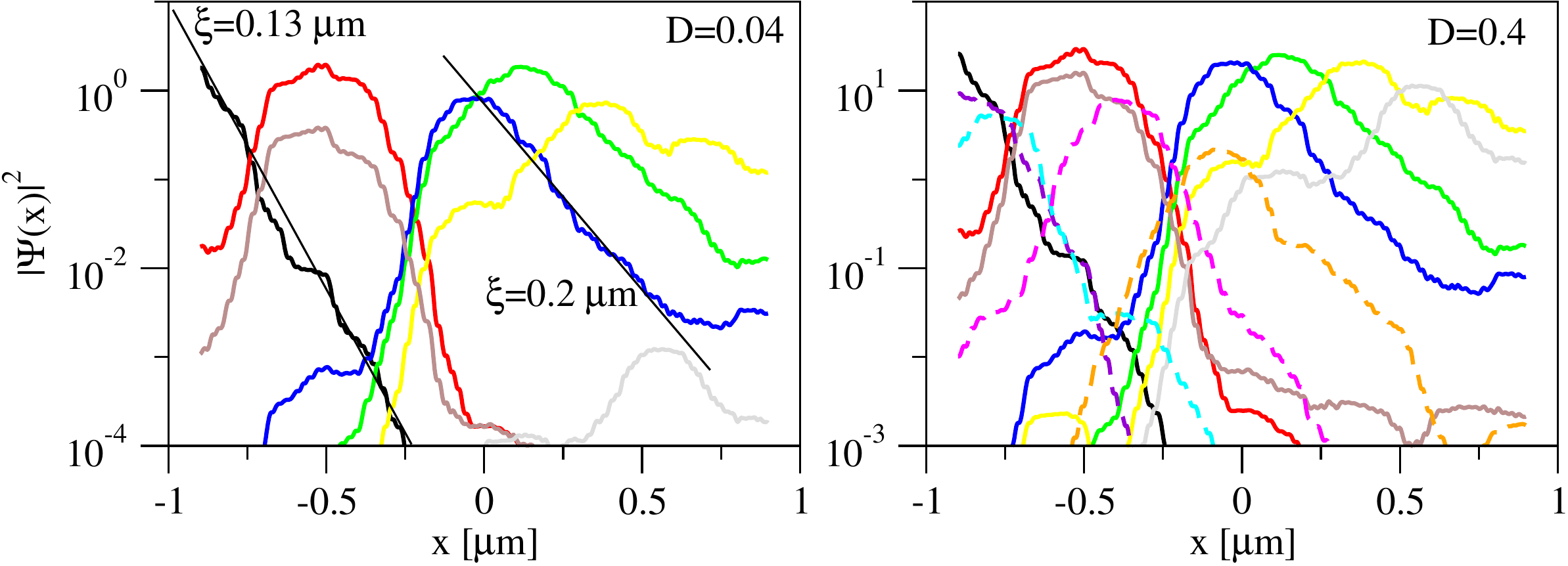}
\caption{
Spatial mode profile $|\Psi(x)|^2$ of lasing modes for a cavity with the same parameters as in Fig.~\ref{fig:localized}. Left panel: at pump strength $D_0=0.04$, there are seven lasing modes. 
Straight lines are exponential fits giving the indicated localization lengths $\xi$. Fits around 
$x=-0.2 \mu $m (not shown) give $\xi \simeq 0.06 \mu $m.
Right panel: at pump strength $D_0=0.4$, three new modes have started to lase (dashed lines),
however, lasing modes have not changed their overall profile, but have only been multiplied
by a mode-dependent factor.}
\label{fig:modes_x}
\end{figure}

We investigate more quantitatively the evolution with pumping of the spatial mode structure.
We show  in Fig.~\ref{fig:modes_x} the spatial profile of lasing modes at two different pump strengths,
both in the multi-mode regime.
At moderate pump strength, $D_0=0.04$ (left panel) there are seven lasing modes
which all exhibit exponential Anderson localization. From exponential fittings, 
we have extracted localization lengths in the range $\xi \in [0.05 , 0.3 ] \mu$m, in good
agreement with the values for the inverse participation ratio shown in Fig.~\ref{fig:localized}a.
Upon increasing the 
pump strength to $D_0=0.4$ (right panel), 
three additional modes are lasing (they are indicated by dashed lines),
however the shape of the lasing modes has not changed at all. Enhanced pumping
only increased their overall magnitude by a mode-dependent factor.

\begin{figure}[t]
\includegraphics[height=5.5cm]{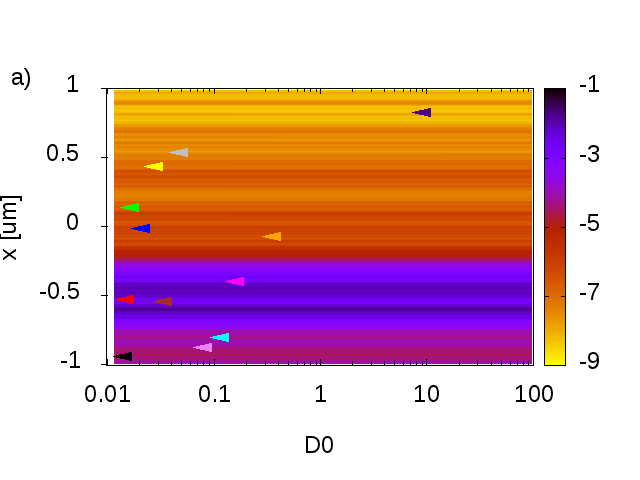}
\includegraphics[height=5.5cm]{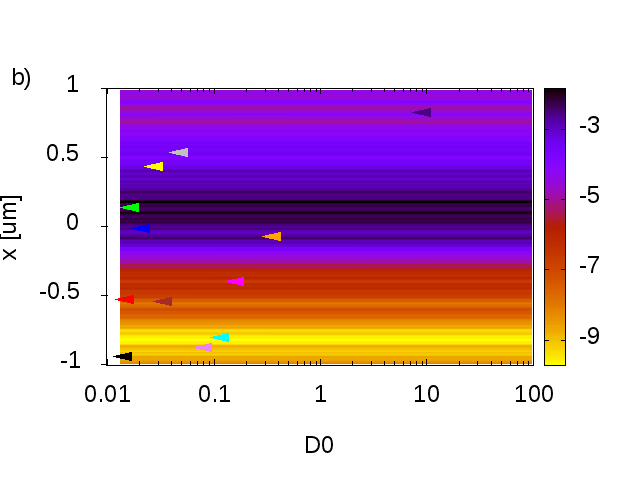}
\caption{Color scale plots of the normalized
spatial mode profile $|\Psi(x)|^2/\int_{\cal C} {\rm d} x |\Psi(x)|^2$ for the second (left) and 
third (right) lasing mode in Fig.~\ref{fig:localized}, corresponding to the 
red and green curves there, respectively.
The arrows show modes thresholds, placed at the centers of mass of the modes, 
with the same color coding as in 
Figs.~\ref{fig:localized} and \ref{fig:modes_x}.
It is clearly seen that the spatial structure
of the modes is insensitive to the pump strength, even for the second lasing mode (left, corresponding
to the red curves in Fig.~\ref{fig:localized}), which strongly overlaps with the brown mode
starting to lase at $D_0 \simeq 0.03$. }
\label{fig:modes_color}
\end{figure}
Fig.~\ref{fig:modes_color} next shows in color scale the normalized 
spatial profile $|\Psi(x)|^2/\int_{\cal C} {\rm d} x |\Psi(x)|^2$ of the second lasing mode,
corresponding to the red curves in Figs.~\ref{fig:localized} and \ref{fig:modes_x}. 
It is clearly seen that over almost four orders of magnitude for $D_0$, and despite
nine mode thresholds being crossed, the mode's spatial structure remains
exactly the same. 
Particularly interesting is the fact that around $D_0=0.3$, a new mode starts
to lase (brown curves in Figs.~\ref{fig:localized} and \ref{fig:modes_x}), spatially right on
top of the mode plotted on the left panel. Despite their strong overlap, these two modes do not
interact, which we attribute to strong localization.
\begin{figure}[t]
\includegraphics[width=18cm]{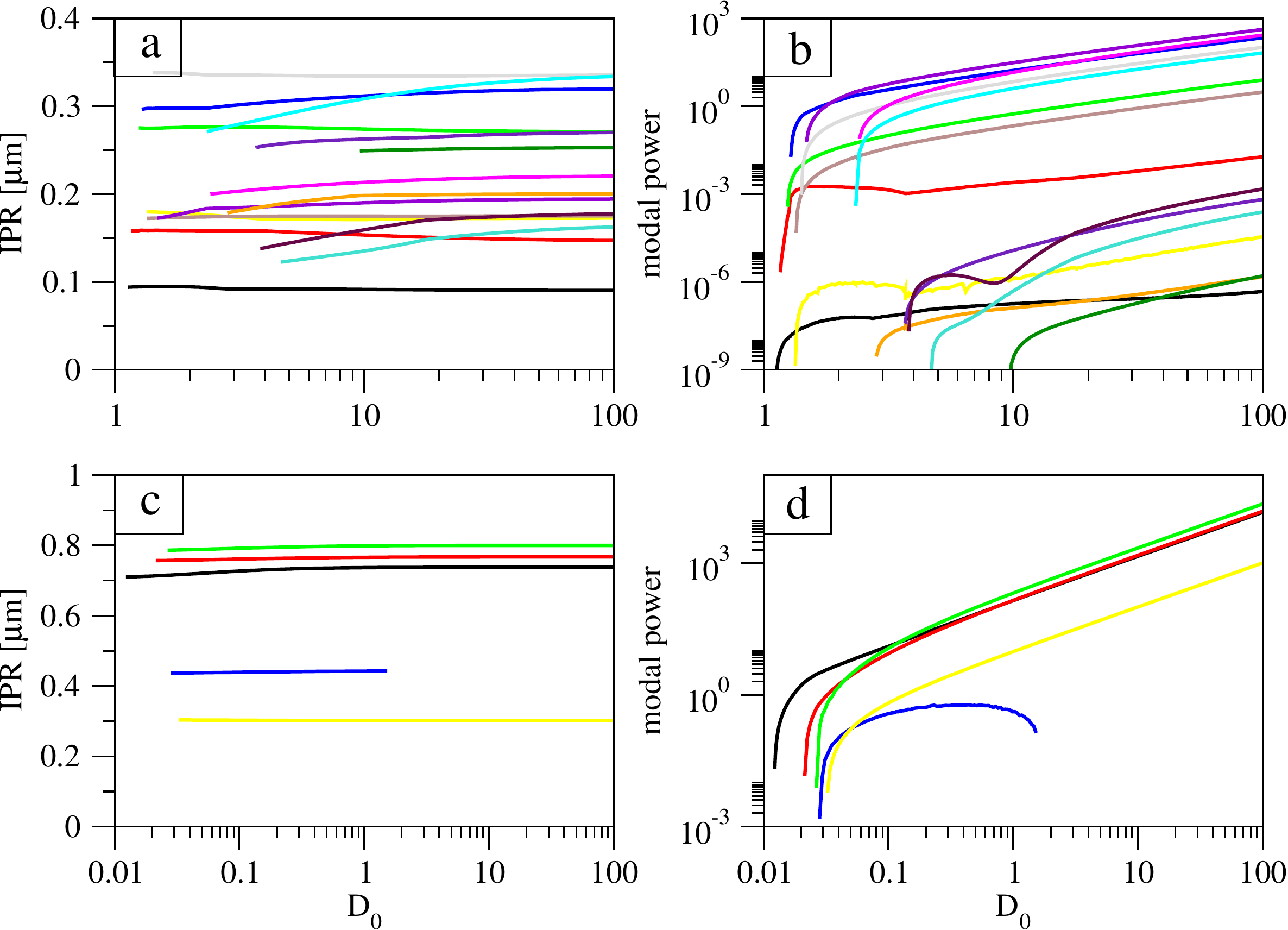} 
\caption{Spatial extent, as measured by the inverse participation ratio (left), and modal
output power (right) of lasing modes, as a function of pump strength $D_0$, for
a non-absorbing, Anderson localized, half-open $1d$ microcavity laser. 
Panels a) and b) are for an absorbing cavity with Re$[n({\bf x})] \in [ 1.3, 6.7 ]$ and  
Im$[n({\bf x})]=0.15$. Panels c) and d) are for a non-absorbing cavity with weaker disorder, 
$n({\bf x}) \in [ 1.3, 4.2 ]$ and extended modes. 
All other parameters are the same as in Fig.~\ref{fig:localized}.}
\label{fig:absorption}
\end{figure}

Having confirmed our prediction that modal interactions are frozen in the Anderson
localized regime, we finally illustrate how they are turned on when the lifetime of the 
passive cavity modes is reduced. To achieve this, we first add some absorption in the cavity
via a finite imaginary part of the index of refraction. We show in Fig.~\ref{fig:absorption}a  and b
data for the same cavity as in Fig.~\ref{fig:localized}, where however the index of refraction
acquires a finite imaginary part, ${\rm Im}[n]=0.15$. 
We see that modal interactions are turned on as inverse participation ratios and modal emission
power vary significantly with the pump strength. We also found, but do not show,
that the lasing frequencies and the centers of mass of the modes vary with the pump strength. 
We finally corroborate our
prediction that modal interactions are turned on in the delocalized regime by simulations
of cleaner cavities in Fig.~\ref{fig:absorption}c and d. We first observe that less modes lase,
which we attribute to the presence of stronger mode interactions. The presence of strong interactions is
further confirmed by the appearance and disappearance of a lasing mode (blue curves) as the pump
increases. Such switching on and off of lasing modes with increasing pump strength have been reported
in two-dimensional weakly disordered cavities~\cite{Tur08}, where they were due to 
frequency repulsion. This is not the mechanism at work here, as we found that the 
blue mode in Fig.~\ref{fig:absorption}c and d is separated
by about $0.5 \gamma_\perp$ from the (yellow) lasing mode with closest frequency.

The interplay between nonlinear interactions and quantum/wave coherent effects in systems with
complex scattering is not well understood.
Here, we have shown how strong localization effects freeze mode interactions in multi-mode lasers,
even between modes that have significant spatial or frequency overlap. Our theory shows that 
Anderson localization justifies a single-pole approximation where modes do not interact, and 
nonlinearities due to spatial-hole burning only determine which modes lase and with what threshold. 
Our results suggest a novel approach to investigate Anderson localization in strongly scattering random lasers 
by observing how much lasing frequencies vary with pump strength, which does not require to
spatially resolve the lasing field.

\section*{Acknowledgements}
We express our gratitude to 
A. Cerjan, L. Ge, A.D. Stone and H. T\"ureci for very helpful discussions on various aspects of
their lasing theory, and to D. Ivanov for discussions on wavefunction correlations with Anderson localization.
This work has been supported by the National Science Foundation 
under grant PHY-1001017.
P.J. acknowledges the support of the Swiss Center of Excellence MANEP and
P.S. acknowlegdes the support of SCIEX.

\newpage

\renewcommand{\thefigure}{S\arabic{figure}}
\renewcommand{\theequation}{S\arabic{equation}}
\setcounter{equation}{0}
\setcounter{figure}{0}





\begin{center}
 {\large \bf Supplementary information}
\end{center}

\section{Steady-state Ab initio Laser Theory}

We give a brief overview of the theory developed by T\"ureci et al.~\cite{Turs} 
for steady-state multi-mode lasing.
Its starting point is the semiclassical
theory of lasing, where the electro-magnetic field is described by the classical
Maxwell's equations, and is coupled to the quantum mechanically
described polarization of the gain medium. We take the latter to be
a collection of two-level atoms. The result is a system of three
nonlinear coupled partial differential equations~\cite{Hakens}
\begin{subequations}\label{eq:mbloch}
\begin{eqnarray}
\partial_t^2 E^+ & = & n({\bf x})^{-2} \left[ \nabla^2 E^+ -4 \pi \partial_t^2 P^+ \right] \, ,\\
\partial_t P^+ & = & -(i \omega_0+\gamma_\perp) P^+ - i g^2 \hbar^{-1}E^+ D \, , \\
\partial_t D & = & \gamma_\parallel (\overline{D}-D) + 2 i \hbar^{-1}[E^+ (P^+)^*-P^+(E^+)^*] \, .
\end{eqnarray}
\end{subequations}
We restrict ourselves to one-dimensional or two-dimensional transverse polarization and accordingly 
consider only scalar field.
In Eqs.~(\ref{eq:mbloch}), the population inversion $D({\bf x},t)$ generates, in the
presence of an electric field $E({\bf x},t)$, a nonlinear polarization
$P({\bf x},t)$ of the atomic medium, which in its turn,
is coupled to $D$ through $E$. 
Here, $\hbar \omega_0$ gives the spacing of
the atomic levels, $\gamma_\parallel$ their relaxation rate and
$\gamma_\perp$ the relaxation rate for the polarization, $\overline{D}({\bf x})$ is the 
pump strength and $n({\bf x})$ is the cavity's index of refraction, related to the dielectric function by $\epsilon=n^2$. Finally, the dipole matrix element 
$g$ couples the electric field with the
population inversion, and we used the standard procedure of writing
the electric field and polarization in terms of their positive
and negative frequency components, $P=P^++P^-$, $E=E^++E^-$, neglecting the coupling of
negative to positive components in the spirit of the rotating wave 
approximation. 

The set of Eqs.~(\ref{eq:mbloch}), augmented by appropriate
boundary conditions, describe nonlinear lasing. 
They cannot be solved exactly, and the standard
procedure is to time-evolve them numerically long enough to
find their steady-state solution. SALT is an alternative which determines
steady-state solutions of 
Eqs.~(\ref{eq:mbloch}) with a much lesser computational price to pay. 
This of course assumes that steady-state solutions exist. We take them as
multiperiodic, i.e. expand the fields $E^+$ and $P^+$
\begin{subequations}
\begin{eqnarray}
E^+(x,t) &=& \sum_\mu  \Psi_\mu({\bf x}) \exp[-i \Omega_\mu t] \, \\ 
P^+(x,t) &=& \sum_\mu  P_\mu({\bf x}) \exp[-i \Omega_\mu t] \, ,
\end{eqnarray}
\end{subequations}
over lasing modes, $\mu = 1,2,... N$.
SALT then uses the stationary inversion approximation, $\partial_t D=0$, 
to determine the components $\Psi_\mu$ of the
expansion as solutions of a self-consistent equation,
\begin{equation}
\left(\nabla^2+n^2({\bf x})k_\mu^2\right) \Psi_\mu({\bf x}) = - k_\mu^2 \frac{\gamma_\perp}{c k_\mu - \omega_0+{\rm i}\gamma_\perp} D_0 \frac{F({\bf x})}{1+h({\bf x})}\Psi_\mu({\bf x}) \, ,
\label{eq:equation for Psi}
\end{equation}
where we parametrized the pump using a volume unit $V_0=g^2/\epsilon_0\hbar\gamma_\perp$ as $\overline{D}({\bf x}) = D_0 F({\bf x})/V_0$, splitting it into an overall dimensionless scale $D_0$ and the spatial modulation $F({\bf x})$, with max$_{\bf x} F({\bf x})=1$.
The non-linearity source (the total weighted electric field intensity, or the hole burning denominator) is
\be
h({\bf x}) = \sum_\nu \Gamma(k_\nu) |\Psi_\nu({\bf x})|^2 E_0^{-2} = \sum_\nu \Gamma(k_\nu) | \sum_p a_\nu^p \psi_p({\bf x},k_\nu)|^2,
\ee
with the Lorentzian weight factors $\Gamma(k) = \gamma_\perp^2 / [ (c k - \omega_0)^2 + \gamma_\perp^2 ]$, and electric intensity units $E_0=\hbar \sqrt{\gamma_\perp \gamma_{||}}/2g$.

Lasing theories usually construct lasing modes
from resonances or quasiresonances of the corresponding (nonlasing) cold 
cavity. The latter are however non-normalizable outside the cavity, so that neither the modal
power nor the relative weight of individual lasing modes can be extracted. SALT instead
expands $\Psi_\mu({\bf x},k_\mu) = E_0 \sum_m a_\mu^m \psi_m({\bf x},k_\mu)$, over a basis of so-called constant-flux $\psi_m$ states which are defined by
\begin{subequations}\label{eq:CF1}
\begin{eqnarray}
\left[ \nabla^2 + n^2({\bf x}) k_{m \mu}^2(k_\mu) \right] \psi_m({\bf x}, k_\mu) = 0,  \;\;\;\;\;\;\;\; {\bf x} \in {\cal C} \, , \label{eq:cavity}\\
\left[\nabla^2 + n_0^2 k_\mu^2 \right] \, 
\psi_m({\bf x}, k_\mu) = 0, \;\;\;\;\;\;\;\; {\bf x} \notin {\cal C} \, ,
\label{eq:free space}
\end{eqnarray}
\label{eq:CF}
\end{subequations}
with the domain ${\cal C}$ defining the amplifying medium, $n_0$ the index of refraction outside and a requirement of containing only outgoing waves at infinity. The latter requirement, together with Eq.~\eqref{eq:free space} can be recast as a condition on the constant flux state at the boundary of a
$d$-sphere enclosing the resonator
($\partial_{\perp}$ is the normal derivative at the surface of that sphere, $d$ is the system's dimension)
\begin{equation}
\partial_{\perp} \psi_m ({\bf x}, k_\mu)= {\rm i} n_0 k_\mu \psi_m ({\bf x}, k_\mu)- \frac{d-1}{2|{\bf x}|} \psi_m({\bf x}, k_\mu).
\label{eq:BC}
\end{equation}
As a consequence of this non-trivial boundary condition,  
the CF basis is not orthogonal in the usual sense. Instead, the following  bi-orthogonality relation holds,
\begin{equation}
\int {\rm d}^d x \, n^2({\bf x}) \overline{\psi}_m^* ({\bf x}, k_\mu) \psi_n ({\bf x}, k_\mu) = \delta_{mn},
\label{eq:ONO}
\end{equation}
between the CF states and their duals $\overline{\psi}_m({\bf x}, k_\mu)$. The latter are defined as solutions of complex conjugates of Eqs.~\eqref{eq:cavity} and \eqref{eq:BC}, which gives $\overline{\psi}_m({\bf x}, k_\mu)=\psi_m^*({\bf x}, k_\mu)$.

Using the ansatz for the electric field intensity $\Psi_\mu$ in Eq.~\eqref{eq:equation for Psi}, one obtains a set of self-consistent equations for the vector of coefficients 
${\bf a}_\mu=(a^1_\mu,a^2_\mu,...)$,
\be
{\bf a}_\mu =  D_0 {\bf T}\, {\bf a}_\mu \equiv D_0 \Lambda \mathcal{T} {\bf a}_\mu \, .
\label{eq:self-consistent}
\ee
Here ${\bf T}$ is the threshold matrix, which depends on the set of vectors $\{{\bf a}_\mu\}_{\mu=1}^N$, and which was split into the following two parts
\begin{subequations}
\begin{eqnarray}
\Lambda_{mn} &=& \delta_{mn} \frac{-k_\mu^2}{k_\mu^2-k_{m\mu}^2}\frac{\gamma_\perp}{ck_\mu-\omega_0+{\rm i}\gamma_\perp},\\
\label{eq:calTmn}
\mathcal{T}_{mn} &=& \int_{\cal C} {\rm d}^d x \frac{F({\bf x})}{1+h({\bf x})}\overline{\psi}_m^*({\bf x},k_\mu) \psi_n({\bf x},k_\mu) \, .
\end{eqnarray}
\end{subequations} 

To calculate the modal power of the lasing state $\mu$, we use a formula derived in Ref.~\cite{Ge2010s}.
With our notation, it reads in $d=1$ dimension
\begin{equation}
P_\mu = 2 \epsilon_0 E_0^2 \omega_\mu S l \int_\mathcal{C} \frac{{\rm d} x}{l} \, |\Psi_\mu(x)/E_0|^2 \left( \Gamma(k_\mu)  D_0\frac{F(x)}{1+h(x)}-{\rm Im}[n^2(x)] \right) \, ,
\end{equation}
where $\epsilon_0$ is the vacuum permeability, and $S$ is the cavity transversal cross-section. 
In the main text, we plot the modal power in dimensionless units, $P_\mu/(2 \epsilon_0 E_0^2 \omega_\mu S l)$,
with $l=53$ nm.

\section{Numerical approach}

Solving Eq.~\eqref{eq:self-consistent} self-consistently is, in general, highly non-trivial. The unknown vectors ${\bf a}_\mu$ enter the threshold matrix elements through the hole burning denominator $h({\bf x})$ [see Eq.~\eqref{eq:calTmn}] effectively leading to an infinite-order non-linearity. In addition, given a pump strength $D_0$, there is no simple way to find out how many modes are lasing, which is a prerequisite information for any numerical algorithm searching for a self-consistent solution to Eq.~\eqref{eq:self-consistent}. To construct the solution, we use an iterative scheme. We first determine
the first lasing mode. Its pump threshold is entirely determined by the passive cavity properties---it is the mode of the largest lifetime in the absence of pumping. From here, the pump is increased by small steps. At each step, Eq.~\eqref{eq:self-consistent} is iterated until convergence is achieved. Finally, the appearance of a real eigenvalue of the threshold matrix surpassing a certain level signals that a new mode starts to lase. This method has been worked out in a series of works by T\"ureci, et al.,\cite{Turs} and has been shown to predict
lasing modes and threshold in complete agreement with the computationally much more expensive
time-evolution of the Maxwell-Bloch equations, Eq.~\eqref{eq:mbloch}.\cite{Ges}

\begin{figure}[t]
\begin{psfrags}
\includegraphics[width=10cm]{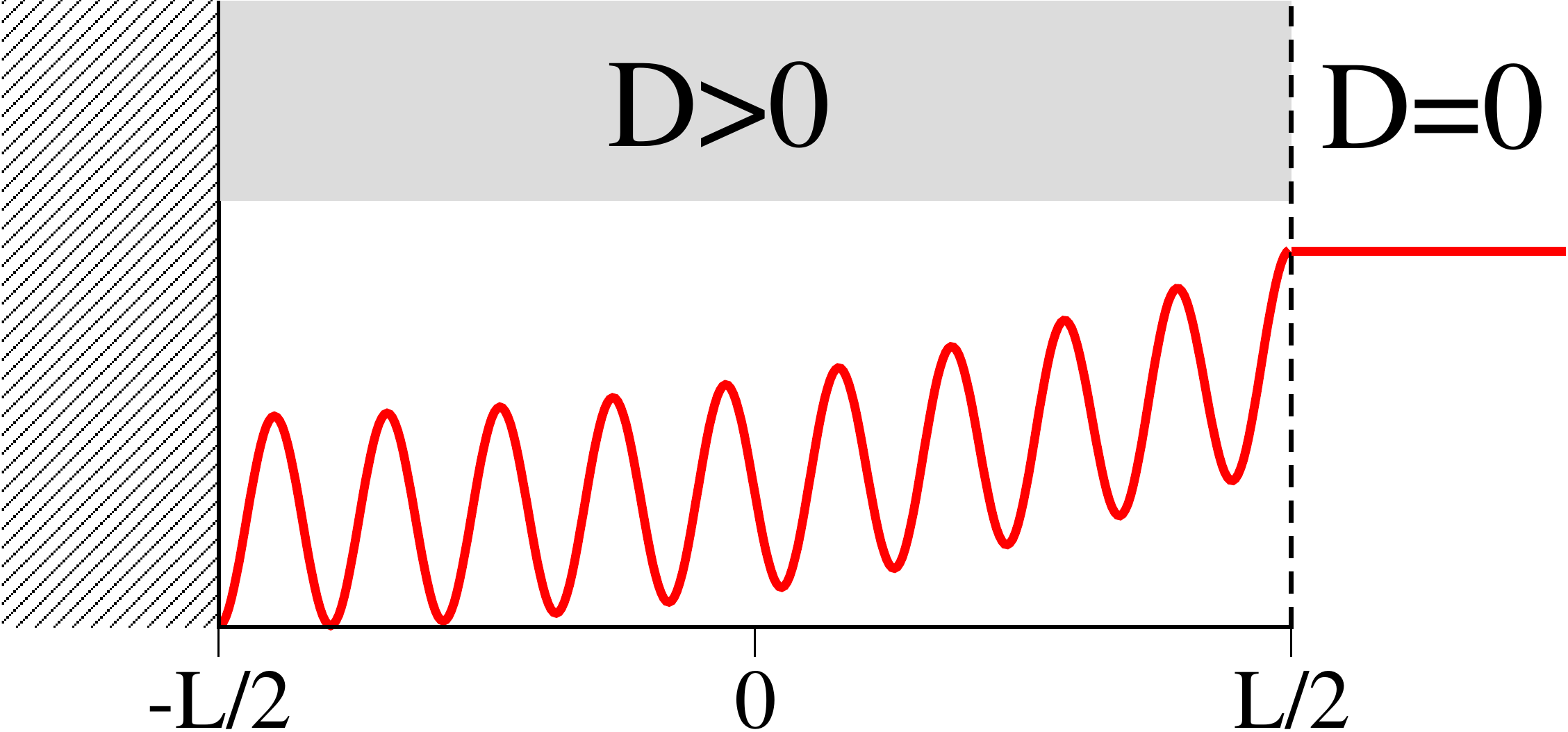}
\end{psfrags}
\caption{Sketch of the microcavity laser we consider, where a $1d$ cavity is closed to the left and
open, hence emitting, to the right (dashed line). The pump area is shadowed. 
The spatial profile of a lasing mode (red) for a uniform pump and a spatially
constant refractive index is plotted for illustrative purpose.}
\label{fig:cavity}
\end{figure}

The microcavity we model in our numerics is depicted in Fig.~\ref{fig:cavity}. It is closed on the left end, open on the right end, and pumped uniformly, $F(x)=1$. In our numerics, we vary the degree of disorder in the refractive index, both in its strength and correlation length $\chi$. The disorder sets the CF states localization length $\xi$. Typical values in our simulations are $\xi\sim100-1000$ nm, $\lambda \sim 50-110$ nm, and $\chi=8$ nm, so that $\xi \gtrsim \lambda \gg \chi$. Here $\lambda=2\pi/n_0k_0$ is the wavelength of a freely propagating light wave in a material with a constant index of refraction, $n(x)=n_0$.

\begin{figure}[t]
\begin{psfrags}
\psfrag{D0}{$D_0$}
\psfrag{x [um]}{$x[\mu m]$}
\includegraphics[width=8cm]{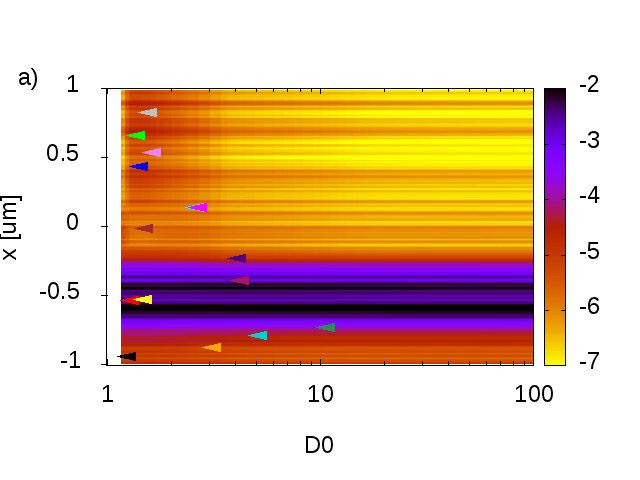}
\includegraphics[width=8cm]{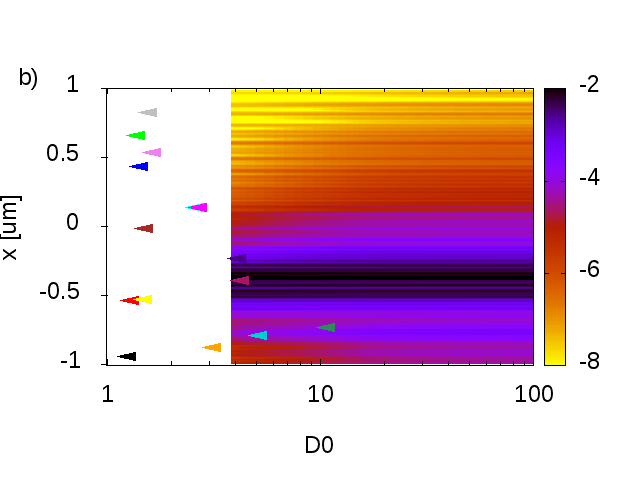}
\end{psfrags}
\caption{Color plots of the 
spatial mode profile ln$|\Psi(x)|^2$ of the 2$^{\rm nd}$ (left) and 13$^{\rm th}$ (right) lasing mode 
for $1d$ edge-emitting, disordered microcavity laser of Figs.~4a-b of the main text 
(corresponding to the red and maroon curves there). Colored arrows indicate the center of mass
positions and pump thresholds of all other lasing modes, with the same color coding as
in Fig.~4a-b.}
\label{fig:modes}
\end{figure}

Fig.~\ref{fig:modes} illustrates structural changes of lasing modes in the interacting case. It should be compared with Fig.~3 of the main text, where a lasing mode is stable even in the exponentially small tails, despite the presence of many other lasing modes. Here we added finite imaginary part to the refractive index, to turn on modal interactions. They result in  clearly visible profile changes of the state plotted in Fig.~\ref{fig:modes}a: after the mode starts to lase, it tends to delocalize (its weight grows in the right half of the cavity). This growth is stopped as soon as other modes are turned on in that part of the space and the plotted mode mainly tends to localize more with enlarging the pump. The opposite behavior towards a weaker localization, is visible on Fig.~\ref{fig:modes}b for another mode. These two examples suggest that there is no generic trend, 
and that 
both localization and delocalization may occur as the consequence of modal interactions.

\section{Structure of the threshold matrix}

The matrix elements of ${\cal T}_{mn}$ given in Eq.~(\ref{eq:calTmn}) are given by the pump profile,
the spatial hole burning denominator and CF wavefunctions. We 
analyze these matrix elements, 
neglecting correlations between
$h({\bf x})$ and $\overline{\psi}_m^*({\bf x},k_\mu) \psi_n({\bf x},k_\mu)=
\psi_m({\bf x},k_\mu) \psi_n({\bf x},k_\mu)$. We evaluate averages
$\langle {\cal T}_{mn} \rangle$ and correlations $\langle {\cal T}_{mn}  {\cal T}_{pq} \rangle$
of these matrix elements over an ensemble of cavities characterized by the same disorder strength
and correlation length, i.e.~the same localization length, but with different microscopic realization of the
disorder (we assume that the pump profile $F({\bf x})$ does not change from one disorder 
realization to the other). In the Anderson localized regime, CF states located deep enough inside
the cavity are long-lived and therefore do not 
differ much from  localized  eigenmodes of closed systems. 
We thus assume that they have the same statistical properties as closed cavity modes. 
Following Ref.~\cite{Mirlins}, we decompose $d=1$ localized wavefunctions as
\begin{equation}\label{eq:psim}
\psi_m(x) = \exp[-|x-x_m|/2 \xi] \, \phi_m(x)/\sqrt{2 \xi} n(x) \, , 
\end{equation}
in terms of an exponential envelope
and a real pseudo-random function $\phi_m(x)$ which 
remains correlated on the scale of the elastic mean free path $\ell$,
$\langle \phi_m(x) \phi_n(x') \rangle = \delta_{mn} \exp[-|x-x'|/2 \xi]$, where
we used the fact that $\ell = \xi$ for $d=1$.

In the Anderson localized regime, the
correlation functions we need are
\begin{subequations}\label{eqs:correls}
\begin{eqnarray}\label{eq:avg}
\langle n^2(x)\psi_m(x,k_\mu) \psi_n(x,k_\mu) \rangle &=& \delta_{mn} e^{-|x-x_m|/\xi} \big/ 2 \xi \, , \\
\label{eq:correl_exch}
\langle n^2(x) n^2(x^\prime)\psi_m(x,k_\mu) \psi_n(x,k_\mu)  \psi_m(x',k_\mu) \psi_n(x',k_\mu)\rangle_{m \ne n} & = &
e^{-(|x-x_m|+|x'-x_m|+|x-x_n|+|x'-x_n|)/2 \xi} \nonumber \\
&& \times e^{-|x-x'|/\xi} \big/4 \xi^2  \, , \qquad
\end{eqnarray}
\end{subequations} 
between eigenfunctions $\psi_n(x,k_\mu)$ of a closed, Anderson localized system
with unbroken time-reversal symmetry.
Here, $x_m$ is the center of mass of $\psi_m(x)$.
In Eq.~(\ref{eq:correl_exch}) we neglect the hybridization contribution coming from even and odd
linear combination of pairs of quasi-degenerate localized states~\cite{Ivanovs} but will comment on 
their possible effect below.
In addition, localized modes have
\begin{eqnarray}\label{eq:disconnect}
\langle |\psi_m(x,k_\mu)|^2 |\psi_n(x',k_\mu)|^2 \rangle & \simeq & \langle |\psi_m(x,k_\mu)|^2 \rangle \times 
\langle |\psi_n( x',k_\mu)|^2 \rangle \, ,
\end{eqnarray}
when they are centered away from one another. 

From Eq.~(\ref{eq:avg}), we first conclude that only diagonal matrix elements ${\cal T}_{mm}$
have finite average,
\begin{eqnarray}\label{eq:9}
\langle {\cal T}_{mn} \rangle = \delta_{mn} \, \int_{\cal C} {\rm d} x \frac{F(x)}{1+h(x)} \frac{1}{n^2(x)}
 \exp[-|x-x_m|/\xi] \big/ 2 \xi \, .
\end{eqnarray}
Unless one specifies $F(x)$ and $h(x)$, this is all one can say. 

We next evaluate
${\rm var} \, {\cal T}_{mn} $. From Eq.~(\ref{eq:correl_exch}) one has
\begin{eqnarray}\label{eq:11}
{\rm var} \, {\cal T}_{mn} &=& \int_{\cal C} \int_{\cal C} {\rm d} x {\rm d} x' 
\frac{F(x)}{1+h(x)} \frac{1}{n^2(x)}
\frac{F(x')}{1+h(x')}  \frac{1}{n^2(x^\prime)}\nonumber \\[3mm]
&& \times \; e^{-(|x-x_m|+|x'-x_m|+|x-x_n|+|x'-x_n|)/2 \xi}  \, e^{-|x-x'|/\xi} \big/4 \xi^2 \, .
\end{eqnarray}
The integral vanishes, unless $|x_m - x_n| \lesssim \xi$, and we substitute
$x_n \rightarrow x_m$ in Eq.~(\ref{eq:11}).
The exponential $\exp[-|x-x'|/\xi]$ in the integrand in Eq.~(\ref{eq:11}) only leads to a prefactor
of order, but smaller than one, since the other exponentials already enforce that 
$x$ and $x'$ are within a localization length away from one another.
We obtain the estimate
\begin{eqnarray}
{\rm var} \,  {\cal T}_{mn}   & \simeq &  \langle {\cal T}_{mm} \rangle^2
 \, ,
\end{eqnarray}
for overlapping modes $m$ and $n$.
Note that ${\cal T}_{mm}>0$ always holds.

We next give an upper bound for the average of 
\begin{equation}
|\Xi|=\left|\frac{ \sum_{n \ne m_0} \Lambda_{nn}  {\cal T}_{m_0 n} {\cal T}_{n m_0} }{
\Lambda_{m_0 m_0}  {\cal T}_{m_0 m_0}^2 } \right| \, ,
\end{equation}
where corrections to the single-pole approximation are justified
when $|\Xi| \ll 1$ [see Eq.~(7) in the main text].
We first note that ${\cal T}_{mn} ={\cal T}_{nm}$ and thus
\begin{eqnarray}
\left|\sum_{n \ne m_0} \Lambda_{nn}  {\cal T}_{m_0 n} {\cal T}_{n m_0} \right|& \le &
{\rm max}_{n \ne m_0} | \Lambda_{nn}| \, \sum_{n \ne m_0}   {\cal T}_{m_0 n}^2 
\le  {\rm max}_{n \ne m_0} | \Lambda_{nn}| \, \sum_{n}   {\cal T}_{m_0 n}^2 \, . \qquad
\end{eqnarray}
We calculate
\begin{eqnarray}
\sum_{n}   {\cal T}_{m_0 n}^2 &=& \sum_n \int {\rm d} x \int {\rm d} x'
\frac{F(x)}{1+h(x)} \, \frac{F(x')}{1+h(x')} \, \psi_{m_0}(x) \psi_{m_0}(x') \psi_{n}(x) \psi_{n}(x') \, \nonumber \\
&=& \int {\rm d} x 
\left[\frac{F(x)}{1+h(x)}\right]^2  \, \frac{\psi^2_{m_0}(x)}{n^2(x)} \, ,
\end{eqnarray}
where we used the completeness relation $\sum_n \psi_n(x) \psi_n(x') = \delta(x-x')/n^2(x)$.
We therefore obtain
\begin{equation}\label{eq:xibound}
|\Xi| \le \frac{{\rm max}_{n \ne m_0} | \Lambda_{nn}|}{|\Lambda_{m_0 m_0}|} \, 
\left( \int {\rm d} x 
\left[\frac{F(x)}{1+h(x)}\right]^2  \, \frac{\psi^2_{m_0}(x)}{n^2(x)} \Big/
\left[ \int {\rm d} x \frac{F(x)}{1+h(x)}  \, \psi^2_{m_0}(x) \right]^2 \right)\, .
\end{equation}
In the Anderson localized regime, the first factor on the right-hand side
is exponentially small, 
${\rm max}_{n \ne m_0} | \Lambda_{nn}| \big/ |\Lambda_{m_0 m_0} = \hbar / \Delta_\xi \tau_\xi \sim \exp[-L/\xi]$,
because (i) $m_0$ has $k_{m_0} \simeq k_{\mu}$ and an exponentially large lifetime
$\tau_\xi \simeq (\xi/c) \exp[L/\xi]$, and (ii) the modes $n \ne m_0$ that 
overlap with $\psi_{m_0}$ differ in frequency by at least $|\omega_n - \omega_{m_0}| 
\gtrsim \Delta_\xi/\hbar$, with the CF mode spacing 
$\Delta_\xi \sim \hbar c/ k_\mu^{d-1} \xi^d$ within one localization
volume. 
The term between parenthesis, when averaged, is an algebraic function of 
$k_\mu \xi$, for any physically relevant choice of $F(x)$, with the form of $\psi_m$
given in Eq.~(\ref{eq:psim}). This is most easily seen, under the assumption
that the pump profile $F(x) = F_0$ is constant over the volume of the sample, which requires
in particular
the active medium to be homogeneous on the scale of the mode wavelength. When this is the
case, we obtain 
\begin{eqnarray}
\int {\rm d} x 
\left[\frac{F(x)}{1+h(x)}\right]^2  \, \frac{\psi^2_{m_0}(x)}{n^2(x)}
&=& F_0^2 \int  {\rm d} x 
 \left[\frac{1}{1+h(x)}\right]^2  \, \frac{\psi^2_{m_0}(x)}{n^2(x)} \nonumber \\
 & \leq &  F_0^2 \int  {\rm d} x 
\frac{1}{1+h(x)} \psi^2_{m_0}(x) \, ,
\end{eqnarray}
so that 
\begin{eqnarray}\label{eq:bound}
\int {\rm d} x 
\left[\frac{F(x)}{1+h(x)}\right]^2  \, \frac{\psi^2_{m_0}(x)}{n^2(x)} \Big/
\left[ \int {\rm d} x \frac{F(x)}{1+h(x)}  \, \psi^2_{m_0}(x) \right]^2 \leq 1 \Big/
\int {\rm d} x \frac{\psi^2_{m_0}(x)}{1+h(x)}  \, .
\end{eqnarray}
We expect this bound to still hold for smoothly fluctuating $F(x)$.
We conclude that $\langle |\Xi| \rangle$ is
bound from above as
$\langle |\Xi| \rangle \lesssim \exp[-L/\xi]$, up to algebraic
corrections. 

This exponentially small upper bound on the average value $\langle |\Xi| \rangle$ does not rule out
that $\Xi$ is large for some disorder realizations. We therefore investigate fluctuations of $\Xi$.
We obtain that they are still given by algebraic functions
of $k_\mu \xi$ times $\exp[-2 L/\xi]$.
The exponentially small factor comes from the prefactor
${\rm max}_{n \ne m_0} | \Lambda_{nn}| \big/ |\Lambda_{m_0 m_0}|$. It can only be 
compensated if ${\cal T}_{m_0 m_0} \rightarrow 0$ as $\exp[-L/\xi]$ or faster. While this is in principle
possible (e.g. for CF modes localized outside of the support of $F(x)$), it never is the case for 
CF modes that contribute to lasing modes. We conclude that $|\Xi| \sim \exp[-L/\xi] \ll 1$ for all 
disorder realizations in the Anderson localized regime. 

In the delocalized regime,
Eq.~(\ref{eq:xibound}) still holds, and the term between parenthesis there
becomes an algebraic function $f(k_\mu \ell,k_\mu L)$ of $k_\mu L$ and $k_\mu \ell$
[Eq.~(\ref{eq:bound}) holds regardless of the
strength of the disorder]. The prefactor ${\rm max}_{n \ne m_0} | \Lambda_{nn}| \big/ |\Lambda_{m_0 m_0}| = \hbar / \Delta_L \tau_{m_0 \mu}$ now depends on the level 
spacing $\Delta_L = \hbar c\big/2 \pi k_\mu L^2$ and the mode lifetime $\tau_{m_0 \mu}$. The latter is
no longer exponentially small.

We close this paragraph with three comments. First, we recall that the infinite
order nonlinearity arising from $h(x)$ generates higher order correlators such as
$\langle \psi_m({\bf x},k_\mu) \psi_n ({\bf x},k_\mu)
[\sum_\nu \Gamma(k_\nu) \sum_{p,p'} a_\nu^p a_\nu^{p'} 
 \psi_p ({\bf x},k_\nu) \psi_p' ({\bf x},k_\nu)]^M \rangle$, which the above treatment neglects,
 and which can be expected to be large for lasing modes. 
 
Second, 
Ref.~\cite{Ivanovs} points out that pairs $(\psi_m,\psi_n)$ of quasi-degenerate localized wavefunctions exist that
are spatially separated by more than $\xi$.
Because of their quasi-degeneracy, even and odd
superpositions $\psi_\pm = (\psi_m \pm \psi_n)/\sqrt{2}$ are quasi-eigenmodes. They give rise to (i)
a slower decay of the correlator in 
Eq.~(\ref{eq:correl_exch}) at small frequency difference $c (k_{m\mu}-k_{n\mu})$ and (ii)
a negative dip in the correlation function of Eq.~(\ref{eq:correl_exch}) a distance $L_M = 2 \xi \,
{\rm ln} [\Delta_\xi/\hbar (c k_{m \mu}-c k_{n \mu})]$. If one such mode lases, the former effect is 
somehow relevant
in that it enhances the ratio ${\rm var} {\cal T}^2_{mn} /
\langle {\cal T}_{mm} \rangle^2$ by a prefactor of order, but larger than one. The emergence of such
Mott states, if they are lasing, can thus lead to an earlier onset of modal interactions. As for (ii)
we believe that this effect is not relevant when $F(x)$ is
extended over more than $L_M$, because $\psi_+$ and $\psi_-$ have equal but opposite correlations at
$x=0$ and $x \simeq L_M$.
This mechanism for hybridization may however turn on modal interactions in Anderson localized random lasers when  $F(x)$ is localized over a region smaller than $L_M$. In particular it may potentially
lead to lasing modes being spatially extended well outside the pump region. However, for such
quasi-degenerate states, the static inversion approximation made to derive the SALT equations
may break down, so that further investigations are necessary to determine the relevance of Mott states in
random lasing in the Anderson localized regime.

Third, we finally stress that from their very nature, all above estimates are parametric
in essence, in that they can be multiplied by numerical prefactors of order one,
depending on the exact form of $F(x)$.

\end{document}